\documentclass[english,aps,pra,reprint,showkeys,noshowpacs,superscriptaddress,floatfix]{revtex4-1}   
\usepackage[T1]{fontenc}	
\usepackage[latin9]{inputenc}	
\usepackage{geometry}		
\usepackage{graphicx}
\usepackage[above,below]{placeins}	
\usepackage{times}
\usepackage{xcolor}
\usepackage{hyperref}  
\hypersetup{colorlinks=true,urlcolor=blue,citecolor=blue,linkcolor=blue}   
\usepackage{natbib}

\begin{document}

\title{Transforming the Preparation of Physics GTAs: Curriculum Development}
\author{Emily \surname{Alicea-Mu\~{n}oz}}\email{ealicea@gatech.edu}
\affiliation{School of Physics, Georgia Institute of Technology, 837 State Street, Atlanta, Georgia 30332, USA}
\author{Carol \surname{Subi\~{n}o Sullivan}}
\affiliation{Center for Teaching and Learning, Georgia Institute of Technology, 266 4th Street NW, Atlanta, Georgia 30332, USA}
\author{Michael F. \surname{Schatz}}
\affiliation{School of Physics, Georgia Institute of Technology, 837 State Street, Atlanta, Georgia 30332, USA}

\begin{abstract}
Graduate Teaching Assistants (GTAs) are key partners in the education of undergraduates. Given the potentially large impact GTAs can have on undergraduate student learning, it is important to provide them with appropriate preparation for teaching. But GTAs are students themselves, and not all of them desire to pursue an academic career. Fully integrating GTA preparation into the professional development of graduate students lowers the barrier to engagement so that all graduate students may benefit from the opportunity to explore teaching and its applications to many potential career paths. In this paper we describe the design and implementation of a GTA Preparation course for first-year Ph.D. students at the Georgia Tech School of Physics. Through a yearly cycle of implementation and revision, guided by the 3P Framework we developed (Pedagogy, Physics, Professional Development), the course has evolved into a robust and comprehensive professional development program that is well-received by physics graduate students. 
\end{abstract}


\maketitle

\section{Introduction}
Graduate Teaching Assistants (GTAs) are essential partners in the education of introductory physics students. In a large-enrollment introductory physics class, students spend nearly as much in-class contact time with GTAs as they do with faculty \cite{gardner}. GTAs typically teach labs or problem-solving sessions (also called ``recitations'' or ``discussions'') to a smaller group of students than faculty do in the lecture (e.g., 20-30 students in a lab/recitation but 100+ in a lecture \cite{maries4}), which means students can get more individualized attention from GTAs than they can from faculty. As a consequence, for many undergraduates, interactions with GTAs could heavily influence their impressions of what it means to be a physicist, and students' attitudes about physics could end up depending more on the GTAs than on the professors \cite{nyquist,bozack}.

The above, combined with the fact that training greatly improves GTAs' confidence, self-efficacy, and pedagogical content knowledge \cite{prieto,luo2,gibbs,baumgartner,bomanthesis,math,dechenne,boman,lin,rosse,leger,dechenne2,reeves1,integrated,ridgway,wheeler2,wheeler,reeves2}, emphasize the importance of providing GTAs with adequate preparation to support their roles as novice educators -- but teaching is not the only thing GTAs do. GTAs are students themselves and there are many demands on their time, such as going to class, doing homework, studying for exams, doing research, attending meetings \cite{bozack,otherthings}, and occasionally to eat and sleep and take a shower. Therefore it is crucial to lower the barrier to engagement in GTA preparation by fully integrating it into the graduate students' professional development.

In this paper we describe the development and implementation of a Physics GTA Preparation course that fully integrates pedagogy, physics, and professional development strategies. Over the years, the course has evolved into a robust and comprehensive program that has been well-received by the GTAs and has had a measurable positive impact on their self-efficacy and approaches to teaching. A companion paper (in preparation) describes the results of our program assessment. We hope that in offering details of the history, development, and content of our curriculum, we can aid other institutions in the planning of their own GTA preparation programs.

\subsection{Institutional Context}
Our GTA Preparation course was designed specifically for first-time GTAs in the School of Physics at the Georgia Institute of Technology, an R1 Institution according to the Carnegie Classification of Institutions of Higher Education \cite{carnegie} and among the top 20\% of physics Ph.D.-granting institutions in terms of graduate enrollment \cite{aipstats}. The graduate student population in the School has varied between 120 and 135 graduate students per academic year in the last five years, and on any given semester roughly 50 of them are employed as GTAs. 

The calculus-based introductory physics classes are required courses for a large majority of Georgia Tech's undergraduate students, a majority of whom are engineering majors. About 1800 undergraduates take these classes in any given semester. The classes have six contact hours per week between undergraduates and instructors: three hours per week of lecture with a faculty member and three hours per week of labs and recitations with a GTA. On any given semester, between nine and thirteen faculty members are assigned to teach the introductory physics lectures and a bit more than half of all the GTAs in the School are assigned to teach two or more of the approximately 70 lab and recitation sections for these classes. 

The vast majority of the GTAs assigned to teach the introductory physics labs and recitations are first-time GTAs, who are also first-year Ph.D. students and the target audience for our GTA preparation course. Each lab or recitation section has an enrollment of around 20-30 students, so a first-time GTA in our School is in charge of anywhere from 40 to 180 undergraduate students, depending on the specifics of their teaching assignment. In addition to teaching labs and recitations, all GTAs for the introductory physics classes are responsible for proctoring and grading exams; additionally, some teaching assignments require the GTAs to grade worksheets and lab reports. The GTAs are not required to host office hours, but they are given the opportunity to do so if they wish. Experienced GTAs, on the other hand, are usually assigned to upper-division or graduate courses, for which they typically grade homework and exams, host office hours, and occasionally guest lecture. Between 2013 and 2020, a total of 174 graduate students have participated in the GTA preparation course, which include the entire population of current graduate students in the School.

\subsection{History and Motivation} \label{history}
There is little available information about GTA training in the Georgia Tech School of Physics before 2010, when the first efforts to provide more preparation to first-time GTAs began. Although no formal preparation program existed at that time, training included these four elements:

\begin{itemize}
    \item \textit{New TA Orientation (NTAO)} -- an Institute-wide half-day meeting before the start of the semester which mostly covered policies (e.g., FERPA, the Family Educational Rights and Privacy Act) and a brief handful of pedagogical topics ($\sim$ 4 hours). 
    \item \textit{Meetings with the GTA Supervisors} -- new GTAs met the coordinators for the introductory physics courses before the start of the semester to go over topics such as proctoring, grading, and general GTA duties and responsibilities ($\sim$ 4 hours).
    \item \textit{Weekly lab/recitation meetings} -- occurring every Friday afternoon during the semester, to discuss the content of the following week's lab or recitation and, in the case of labs, set up any necessary equipment ($\sim$ 1 hour per week).
    \item \textit{Pedagogy Seminars} -- run by the Georgia Tech Center for Teaching and Learning (CTL) in the first two months of the semester ($\sim$ 5 hours).
\end{itemize}

There were several problems with this piecewise approach to GTA training. First and foremost was the complete disconnect between pedagogy and physics content. The GTAs learned a few basics of pedagogy in very general contexts, with little to no connection to physics in general nor to the specific physics content they would be teaching. At the same time, the physics content training focused almost exclusively on troubleshooting equipment, with conceptual understanding only covered on a need-to-know basis and with zero pedagogical backing. Another problem was the lack of pedagogical reinforcement -- whatever little pedagogy GTAs learned during the NTAO would never again be revisited during the semester. Theoretically, the CTL seminars should have provided such reinforcement, but in practice this was not the case. GTAs were quite vocally unhappy about the scheduling of these seminars (6pm on Fridays) and many appeared to strongly resent being taught how to teach by someone who was not a ``physics person.'' The pedagogy training was thus essentially outsourced, something that the research literature deems as not ideal for STEM GTAs \cite{luft,harris,ellis,TAworkshop}, as this can leave them with the impression that pedagogical knowledge is not relevant or important to their actual teaching duties.

In addition to these problems, the absence of a coherent and unified GTA training program meant a lack of mentoring and career development opportunities, resulting in many unmotivated GTAs who seemed to think about teaching as a burden they must get through in order to get paid instead of as an essential aspect of their development as physicists.

\subsection{The Pilot Semester} \label{pilot}
The Center for Teaching and Learning at Georgia Tech began a ``Super TA'' program in 2012 through which they integrated GTA training into the academic units while still coordinating from a central entity \cite{tris,carol}. The School of Physics joined this effort in 2013.

A Super TA is an experienced GTA who is further trained in pedagogy by taking teaching and learning courses offered by CTL, and who works in conjunction with a CTL mentor to adjust the standard GTA training course into something specifically useful for their particular discipline. In Fall 2013 we developed and deployed the pilot semester of the Physics GTA Preparation course, with one of the authors (EAM) as the Super TA.

The course consisted of a day-and-a-half \textit{Orientation} before the start of the semester (topics covered: Active Learning, Creating Engaging Explanations, Professionalism, Georgia Tech Policies, Time Management, Microteaching, Classroom Management), plus five \textit{Follow-Up Meetings} spread out during the semester (topics covered: Group Work, Grading, Leading Discussions, Midterm Evaluations, Writing a Teaching Philosophy Statement). The lessons were taught workshop-style, with discussions and activities designed to engage the GTAs, thus modeling for them the type of instruction we expected them to implement when teaching their own classes. Additionally, we presented various teaching scenarios in the form of case studies, which prompt GTAs to think about and reflect on what they would do if they are presented with a similar situation in the classroom \cite{casestudies}.

While teaching the class, we were able to assess the relative success of each lesson via informal observations of how engaged the GTAs were in each class meeting. For example, the first day of the Orientation started at 9am and ended at 5pm. We could tell the GTAs were restless and grumpy and wanted to leave by around 3 o'clock in the afternoon. Some GTAs reacted with reluctance, even combativeness, towards the very concept of active learning, something that other researchers have also observed \cite[e.g.,][]{riot1}. GTAs were dismissive about the topic of leading discussions because they claimed no such thing would ever come up in a physics lab or recitation. Similarly, many GTAs said that writing a teaching philosophy was not really useful for their professional development because they did not plan on staying in academia after graduation. More concrete information came from the final reflection assignment, where GTAs elaborated on which course topics had the most impact on them. The three topics most frequently mentioned that first year of the course were Microteaching, Grading, and Midterm Evaluations. From this we could determine that GTAs in general preferred material that was practical and directly applicable to their teaching, and disliked heavily theoretical topics or topics that did not appear relevant for their future careers.

These observations led to a substantial investment of time and effort to modify and enhance the curriculum for the Physics GTA Preparation course, so the class could provide the GTAs with better preparation for their teaching duties and give them opportunities for professional development.

\section{Curriculum Development}\label{currdev}

\subsection{Theoretical Background}
\begin{figure}
    \begin{center}
	\includegraphics[width=0.4\textwidth]{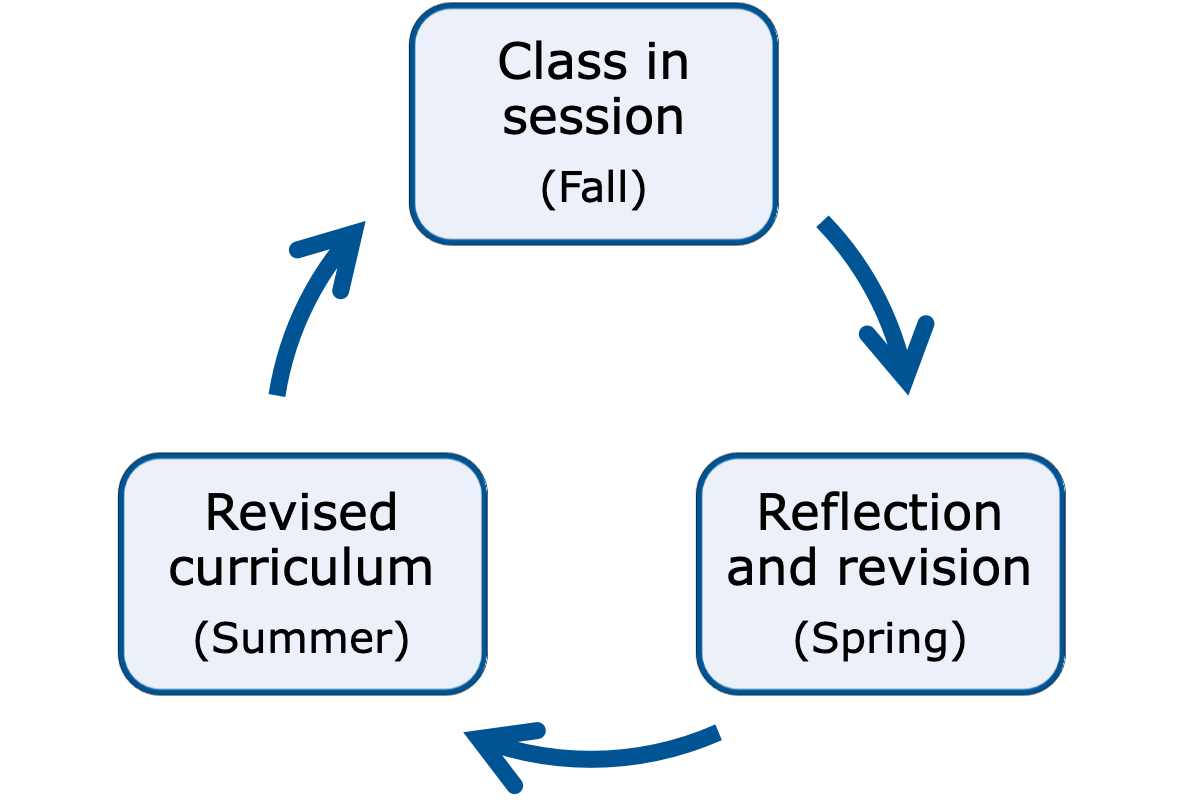}
	\caption[Cycle of implementation/revision of the Physics GTA Preparation class]{Cycle of implementation and revision of the Physics GTA Preparation class. Figure adapted from \cite{myAPStalk}.}
	\label{thecycle}
	\end{center}
\end{figure}

The Fall 2013 pilot followed the course design described by \cite{tris} and \cite{carol}. The initial design was informed in part by process education, an educational philosophy that focuses on the development of broad, transferable learning skills \cite{process}. Changes to the curriculum have been grounded in the principles of instructional design \cite{fink}, and are worked through in a cycle of implementation and revision (see Figure~\ref{thecycle}).

\begin{itemize}
    \item The class is offered in the Fall semester to all first-year Ph.D. students who are concurrently enrolled in a Graduate Teaching Assistantship. Various assessments are carried out throughout the semester when the class is in session (details to be provided in a companion paper, in preparation). The instructor writes down her observations of GTA engagement and discusses them with her CTL mentor and the GTA supervisors.
    \item In the following Spring semester we compile the assessments and the instructor's notes, and plan for necessary changes and revisions to the curriculum. We take into consideration the GTAs' feedback provided to the instructor in terms of what aspects they found useful and would like more of, and what elements of the course need improvement. We seek additional information from the GTA supervisors in case there are changes to the teaching assignments that need to be incorporated into the GTA preparation class.
    \item The revised curriculum (e.g., class materials, slides, activities) is crafted during the Summer term. At the end of Summer, the revised curriculum is implemented and the cycle begins anew.
\end{itemize}

The implementation and revision process can be contextualized in the light of two similar theoretical frameworks: \textit{Design Research} and \textit{Action Research}. Design Research (DR) investigates how people function in a real learning environment by designing experiments with successive refinement based on data \cite{design2,design3,design1}. Action Research (AR) is a recursive, reflexive, dialectical technique used to iteratively revise plans and implementations of educational reforms \cite{bodner3,hunter,bodner}. These two frameworks share many epistemological, ontological, and methodological underpinnings, and have a common meta-paradigm -- pragmatism \cite{dbr}. In terms of both DR and AR, feedback from the GTAs is crucial for the revision part of the process \cite{jones,bovill}.

The course activities were created with constructivism \cite{bodner2,little} and active learning \cite{transforming,scientificapproach} in mind, so GTAs learn how to teach in the same way that they are expected to teach \cite{mutambuki}. The course experience as a whole can be framed within situated learning theory, an approach based on constructivist epistemology in which learners construct new knowledge by connecting prior experience to active participation within a community of practice \cite{lave,socialization2,wheeler3,milner}. The community of practice emerges organically; all the GTAs in the class share the experience of teaching at Georgia Tech for the first time (and of being first-year Ph.D. students taking the core graduate physics courses), and participation in this class gives them a sense of everyone being ``in it together'' \cite{holmes}.

Course content and design follow the best practices in GTA preparation that can be found in the research literature \cite{mythesis}. The course is designed as a meaningful partnership between graduate students and faculty \cite{rhodes,meyers,alan,goertzenthesis,kendall,linenberger}, and spreading the sessions throughout the Fall semester ensures the training is an ongoing endeavor \cite{maries4,mutambuki,shannon,etkina,roehrig,hume,verley}. The GTAs are given the opportunity to practice, and they are observed in action and given constructive feedback \cite{jones,holmes,shannon,etkina,videotape,black,goertzen2,thomas,scaleup,spfin2,gormally,taiop,parker,hickok,telephone}. The content is grounded in research-based teaching strategies \cite{mutambuki,park,pinder}, and takes into account GTAs' beliefs in order to foster a sense of professional identity and buy-in for reformed teaching \cite{goertzen2,goertzen1,ebertmay2,brownell,sandi1,ebertmay,spfin1,flaherty,gretton,lee}. Finally, the course highlights the transferable skills that GTAs will be able to use in their post-graduate school life regardless of their chosen career path \cite{chism,keyskills,jobskills,holaday,altcareers}.

\subsection{The 3P Framework}
The main goals of our physics GTA preparation course are to produce GTAs who are motivated and effective teachers, and to help GTAs develop professional transferable skills that can be used outside of the classroom. The first round of curriculum revision resulted in a specification of the course's learning objectives to achieve the expected course goals:

\begin{enumerate}
    \item Developing and applying learner-centered teaching practices to create a valuable, student-centered, learning experience
    \item Explaining physics concepts, addressing students' preconceptions, and facilitating problem solving
    \item Applying teaching principles to giving and receiving feedback to revise and improve their teaching practice
    \item Managing classroom dynamics and developing efficient ways of grading students' work
    \item Reflecting on their professional identity and identifying transferable skills utilized in teaching that are useful for their future careers as professional physicists
\end{enumerate}

We identified three major themes in these objectives: (1) Pedagogy, (2) Physics, and (3) Professional Development. In the process of trying to determine what themes were served by each of the course topics in the Fall 2013 pilot, we realized that several items could fit into one or more of the main three themes. 

\begin{figure}
    \begin{center}
	\includegraphics[width=0.4\textwidth]{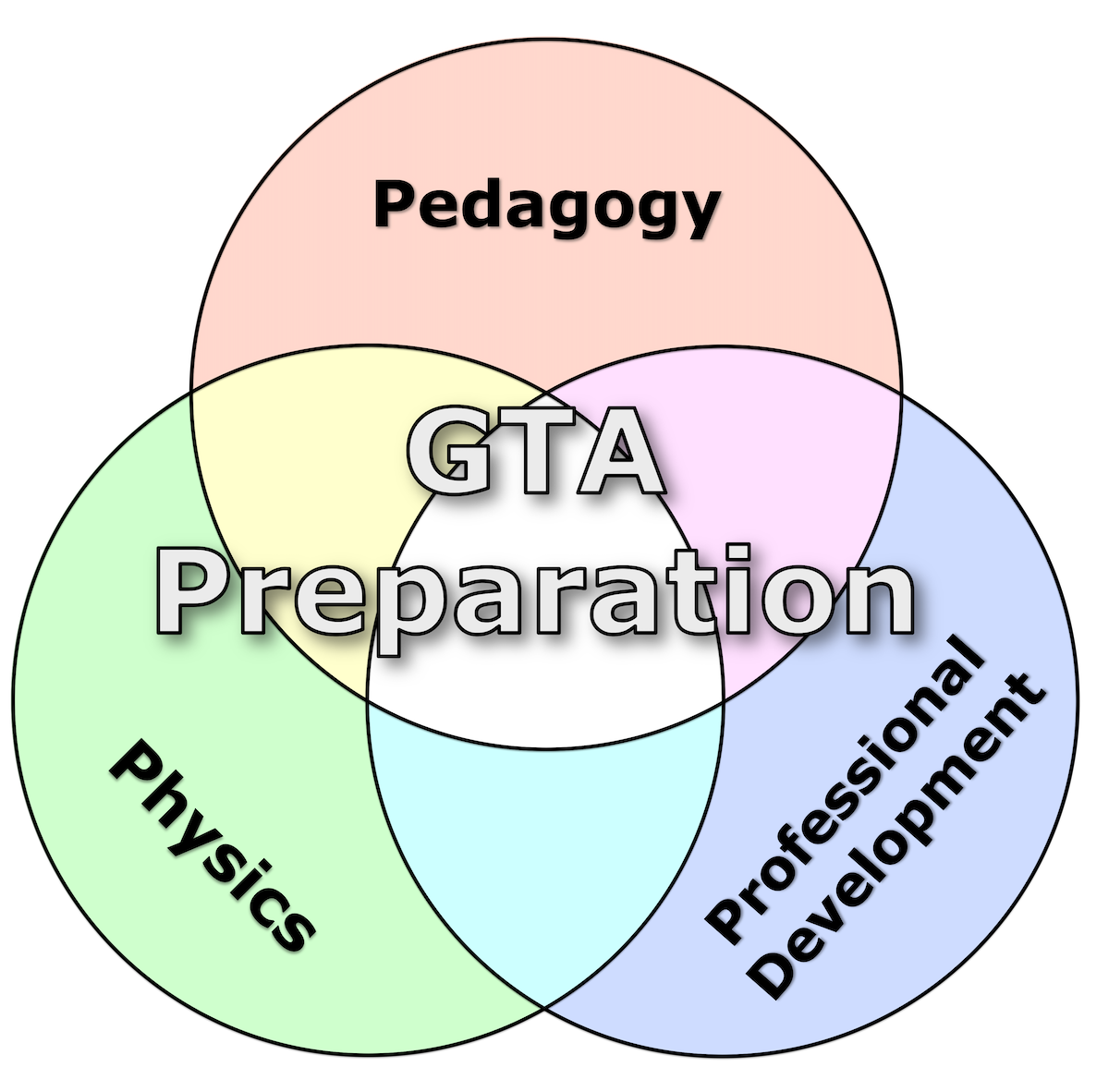}
	\caption[The 3P Framework]{Illustration of the 3P Framework, and how the integration of Pedagogy, Physics, and Professional Development leads to a comprehensive GTA Preparation program.}
	\label{venn}
	\end{center}
\end{figure}

In order to make curricular improvements we developed a method for integrating these three main themes into what we are calling the \textbf{3P Framework}. This framework posits that in order to have a comprehensive program for GTA preparation that is useful and valuable for GTAs in the classroom and beyond there must be full integration between pedagogical knowledge, physics content, and professional development strategies. Pedagogy alone is not enough, because GTAs are novice instructors who need guidance in how to apply pedagogical knowledge to their physics teaching assignments. Physics alone is not enough, because just knowing the content does not guarantee skills in how to teach it. And professional development is crucial for motivation, so GTAs will see that teaching can help them achieve their professional goals even if they lie outside of academia. The 3P Framework can be visualized with a Venn Diagram (Figure~\ref{venn}) in which each circle corresponds to one of the P's: Pedagogy, Physics, Professional development.

A key feature of the 3P Framework is that the intersections of the three P's are just as important as each of the three P's themselves. For example, the intersection of Pedagogy and Physics (and a good chunk of the Physics section itself) is inspired by the Pedagogical Content Knowledge (PCK) framework, which is defined as ``the particular form of content knowledge that embodies the aspects of content most germane to its teachability'' \cite{pck}. PCK goes beyond knowing the content, since it requires the teacher to understand what makes certain topics easy or difficult for their students to learn. It is acquired by reading education literature, watching experienced teachers, and by teaching and reflecting on one's own practice \cite{grayson}. Although the PCK framework was developed with K-12 teachers in mind, and there are some notable differences in the needs and beliefs of novice teachers and GTAs \cite{gormally,gardner}, we find that it is a useful foundation for the intersection of Physics and Pedagogy. GTAs need to develop an understanding of their students' prior knowledge, preconceptions, and mindset for learning \cite{pellathy,posner,aguilar}, but as novice teachers they lack a theoretical background in education which means it is not enough for them to know physics to be effective physics instructors \cite{singh1,singh2,seung1,seung2}. This is especially important within the context of interactive engagement, where GTAs must be able to anticipate, engage with, and build upon student ideas in the classroom \cite{spfin2,spikethesis}.

\subsection{Mapping the Curriculum}
After the pilot, we mapped the course contents onto the 3P Framework. This mapping can be found in the left panel of Figure~\ref{firstlast}. The reasoning behind this mapping is as follows:

\begin{figure*}
    \begin{center}
	\includegraphics[width=\textwidth]{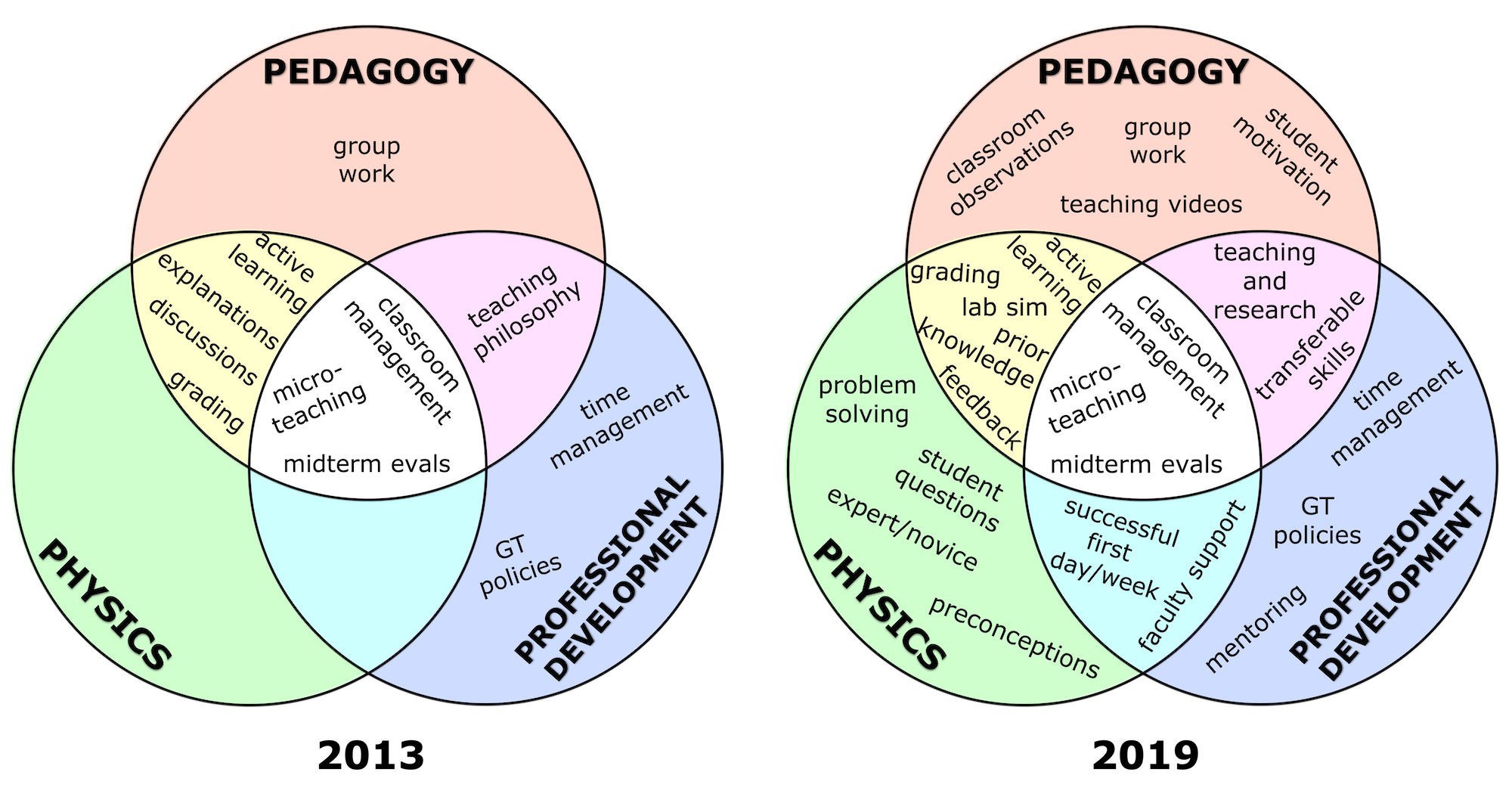}
	\caption[Curriculum mapping according to the 3P Framework]{Mapping the course contents onto the 3P Framework. Left panel: the pilot semester (Fall 2013). Right panel: the most recent in-person semester (Fall 2019). Note that the gaps present at the beginning are now filled.}
	\label{firstlast}
	\end{center}
\end{figure*}

\begin{itemize}
    \item The Group Work lesson did not include anything specific about physics, so we considered it to be pedagogy.
    \item The Georgia Tech Policies and Time Management lesson had nothing to do with physics or pedagogy, but understanding them both can help GTAs in their professional development.
    \item Writing a Teaching Philosophy contributes towards the professional development of GTAs but likely only if they stay in academia, in a teaching-focused position.
    \item Active Learning, Creating Engaging Explanations, Leading Discussions, and Grading lessons were supposed to be pedagogical topics applied in a physics context, although in practice, the physics content for these lessons was sparse in the pilot semester.
    \item We determined that Microteaching, Midterm Evaluations, and Classroom Management lessons included aspects of all three P's, so we placed them in the center of the diagram.
\end{itemize}

Mapping the course contents this way revealed a large gap in the Physics aspect of the framework, thus indicating that the course was far from comprehensive. At that time we began the cycle of revision and implementation. The right panel in Figure~\ref{firstlast} represents the mapping of course topics in the most recent in-person iteration of the course, Fall 2019. We can see that in the most recent version there are no gaps, thus the course is now robust and comprehensive. It should be noted that the main structure of the class (Orientation and Follow-Up Meetings) remains unaltered, but with some additional out-of-class activities. 

Under the 3P Framework, our GTA preparation course has evolved from ``pedagogy and logistics with sparse physics content'' into a robust and comprehensive professional development program that is well-received by the GTAs, is considered useful for their first teaching assignment, and that highlights the ways in which teaching can help them hone their transferable professional skills. As such, we propose that the 3P Framework can be applied to other institutional or disciplinary contexts to guide the development of a similarly robust curriculum.

\subsection{Time Commitment}
The piecewise training that existed prior to the development of the GTA preparation course amounted to approximately 13 contact hours for the GTAs (see Section~\ref{history}), outside of the weekly lab/recitation prep meetings. The pilot semester of the course (Fall 2013) had a total of 17 contact hours. As expected, adding content and activities to make the program more robust resulted in an increase in the time commitment, though not a large one -- the content and delivery has been streamlined so that the Fall 2019 course had a total of 20 contact hours, only a three-hour increase from the pilot semester.

Although these 20 hours do mean an additional workload for first-year physics Ph.D. students, who are already quite busy with teaching and taking classes, we consider it an essential investment for their professional development. Without this course, first-time GTAs would fall back into a piecewise training approach which would still result in a not-insignificant time commitment while providing them with vastly inadequate support. The other alternative of providing no training at all should be out of the question, not just because of the benefits to GTAs' confidence, self-efficacy, and pedagogical content knowledge, but also because the undergraduates enrolled in introductory physics labs and recitations benefit from having GTAs who who have been taught how to teach.

We have also explicitly asked the GTAs, in end-of-semester surveys, whether the time commitment for the course seems reasonable or excessive. A majority of them have said the course is not a burden and the time they spend on it is reasonable. They also appreciate being able to have lively discussions with their peers, which is not something they get to do in their other classes.

\subsection{Sustainability}
The success of our GTA preparation course would not have been possible without the commitment and support from the departmental administration and our partnership with the Center for Teaching and Learning (CTL), who sponsor activities such as a lunch for new and experienced GTAs. Since other departments also partner with CTL for their own GTA preparation courses, we are able to have discussions with other GTA developers across campus and share ideas and resources.

The course was at first taught by a graduate student whose teaching assignment was specifically for the GTA preparation course (the ``Super TA'' mentioned in Section~\ref{pilot}). The course now has a dedicated non-tenure-track faculty member as the instructor (the same Super TA after graduating), whose teaching duties explicitly include preparing GTAs for teaching. When possible, an experienced GTA is assigned to the class to help with activities such as classroom observations. 

In terms of materials and equipment, the only thing that was necessary in addition to regular office supplies was video recording equipment. A modest expense of approximately \$2000 allowed us to purchase two sets of video cameras and microphones for use in classroom observations.

\section{Course Structure and Content}
The Physics GTA Preparation course is structured in two parts: the \textit{Orientation} and the \textit{Follow-Up Meetings}. The Orientation comprises roughly 3/4 of the total contact hours of the class and happens before the semester begins. The Follow-Up Meetings happen every 2-3 weeks during the semester. The course also includes out-of-class work and activities such as classroom observations, workload surveys, and mentoring meetings. Table~\ref{c2019} shows the structure of the course in Fall 2019, which will be described in detail in this section. A full set of course materials can be found at \url{https://tinyurl.com/ealiceaGTAPD}.

\begin{table*}
\caption{\label{c2019}Physics GTA Preparation course structure (2019).}
\begin{ruledtabular}
\begin{tabular}{p{0.2\linewidth} p{0.75\linewidth}}
Module & Brief Description\\
\hline
\multicolumn{2}{c}{\textit{Orientation}}\\
Intro \& GT Policies & GTA duties and expectations; Georgia Tech Policies\\
Teaching Physics & Active learning; explaining concepts and addressing preconceptions; the novice/expert divide and anticipating student questions; facilitating problem-solving\\
Classroom Management & Strategies for classroom management; facilitation of group work; how to keep students motivated\\
Lab Simulation & Practice teaching in a lab environment using real introductory physics lab experiments\\
Microteaching & Practice teaching problem-solving, and giving and receiving feedback from peers and instructors\\
\hline
\multicolumn{2}{c}{\textit{Follow-Up Meetings}}\\
Grading & Strategies for fair and efficient grading, including rubrics; grading practice with old exam problems\\
Midterm Evaluations and Time Management & Strategies for collecting teaching feedback from students; strategies for effectively managing the time spent on different tasks\\
Teaching Videos & Watch and critique video recordings of past physics GTAs at Georgia Tech\\
Teaching and Research & Identifying transferable skills in teaching that can help in future careers beyond the classroom\\
Concluding Remarks & Final thoughts and reflections at the end of the first semester of graduate school\\
\hline
\multicolumn{2}{c}{\textit{Out-of-Class Activities}}\\
Workload Surveys & Weekly surveys to indicate the time spent on various GTA and graduate student duties\\
Classroom Observations & An instructor observes each GTA twice per semester and provides them with feedback\\
Mentoring Meetings & Peer mentoring sessions with senior graduate students\\
\end{tabular}
\end{ruledtabular}
\end{table*}

\subsection{Orientation}
The Orientation is the first part of the GTA preparation class. Each session is three hours long, and they are spread out over a period of several days on the week before the semester begins and the first week of the semester (GTA duties begin on the second week of the semester).

\subsubsection{Introduction \& Georgia Tech Policies}\label{introGTp}
This three-hour module is the first meeting with the new graduate students. The lesson is structured into four parts: (1) \textit{Introductions and Syllabus}, where the course instructor and the GTAs introduce themselves, followed by a discussion of the course syllabus (schedule, requirements, assignments, grading scale); (2) \textit{GTA Duties and Expectations}, where we explore what the GTAs' expectations are for their first semester of graduate school and for their teaching experience, and discuss the duties and responsibilities of each GTA assignment; (3) \textit{Georgia Tech Policies}, a discussion of Georgia Tech's Policy of Nondiscrimination, Academic Integrity, the Office of Disability Services (ODS), Sexual Misconduct, and FERPA; (4) \textit{OK/NOT-OK Game}, in which GTAs are presented with short scenarios related to the previously discussed Georgia Tech policies, and they say whether each presented scenario is acceptable (``OK'') or unacceptable (``NOT-OK'').

The lesson ends with a prompt for the next day, asking GTAs to think about what are the best ways to teach and learn physics. GTAs are also given the opportunity to write down what, if anything, is still unclear after this lesson, and the questions are answered in the following class meeting.

\subsubsection{Teaching Physics}\label{js2}
The second three-hour module focuses on pedagogical content knowledge for teaching physics. It starts by asking the GTAs to express their thoughts about the previous day's ending prompt -- the best ways to teach physics and the best ways to learn physics. It then flows into a discussion of differences between experts and novices \cite{chipaper}, and how to ``unpack'' students' questions. A couple of videos are shown of GTAs interacting with students who have questions; each video is followed by a discussion of the good and not-so-good things in the video. The videos feature previous GTAs in the Georgia Tech School of Physics.

The discussion about questions from students leads into a discussion about asking students questions, emphasizing the need to connect new knowledge to the students' prior knowledge. We then move on to talk about incorrect prior knowledge, in particular preconceptions and misconceptions. We introduce concept inventories, where the incorrect answers can help us identify students' preconceptions, and do an activity titled ``Identifying Misconceptions, or Why did they get it wrong?'' The GTAs are separated into groups, and each group is assigned one pre-selected problem from a concept inventory. The groups then discuss reasons why students would pick the incorrect answers and how they would address each of those misconceptions. When the activity is over, everyone takes a break.

After the break we begin with an introduction to active learning, emphasizing its effectiveness and contextualizing it within PER \cite[e.g.][]{hake}. We then discuss problem solving, and how experts and novices differ in their approaches to solving physics problems. In the final activity of the session, GTAs are once again separated into groups and each group is given one introductory physics problem to solve and to identify the types of issues that students may have when attempting to solve each of the problems. The lesson ends by introducing the Microteaching activity, with each GTA selecting the problem they will microteach. The handout packet associated with this lesson includes excerpts and adaptations from the book ``Five Easy Lessons: Strategies for Successful Physics Teaching'' \cite{fiveeasylessons}, and lists all references used so GTAs may consult them directly if they so wish.

\subsubsection{Classroom Management}
This lesson starts with a discussion of the first day of teaching -- what they need to do and what they want to accomplish. We emphasize the importance of establishing credibility, and the necessity of setting expectations for the students \cite{gaffney}.

A discussion of two case studies follows: the first is an example of what not to do if there are classroom incivilities, and the second is a worst case scenario for a lab in which too many students require the GTA's attention at the same time. Case studies are a great way for GTAs to ``experience'' a teaching situation, and encourages them to reflect on the decisions they would make when presented with different scenarios \cite{casestudies}.

The case studies then flow into a discussion of how to efficiently facilitate group work, with video examples of GTAs at work in various contexts. Most videos feature past GTAs in the Georgia Tech School of Physics, but we have also used video clips from Periscope \footnote{Periscope, \url{https://www.physport.org/periscope}}\cite{periscope}. 

The last part of the lesson focuses on student motivation as a function of three dimensions: self-efficacy, value, and environment \cite{howlearningworks}. Additional case studies are then presented for discussion. It should be noted that all the case studies covered in this lesson are based on true stories that the course instructor has experienced, observed, or heard about.

\begin{figure*}[ht]
    \begin{center}
	\includegraphics[width=\textwidth]{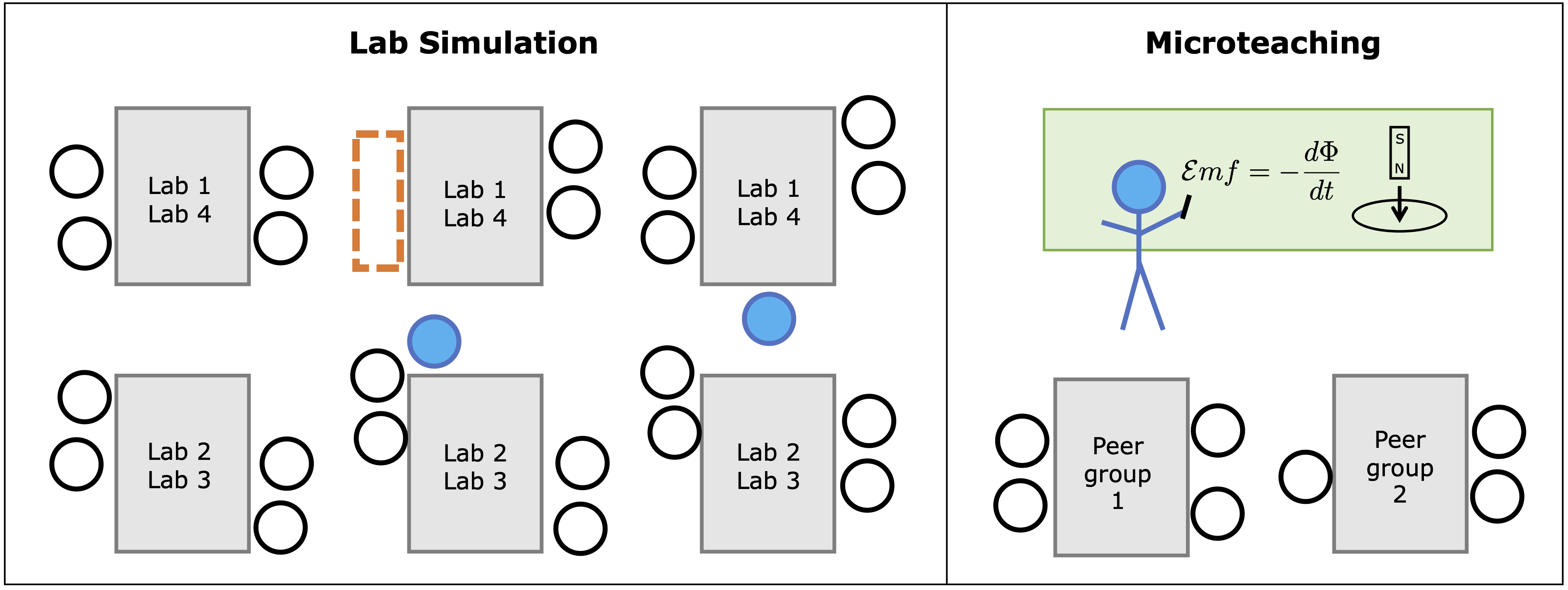}
	\caption{Schematics of the classroom settings for the Lab Simulation and Microteaching activities. Left panel: top view of the Lab Simulation classroom. Each gray rectangle represents a lab table, the empty circles represent the participants acting as Students, the filled blue circles represent the two participants acting as TAs, and the orange dashed rectangle indicates where those two participants will sit when it is their turn to be Students. Right panel: each Microteaching session has GTAs taking turns to facilitate a problem solving session, while the rest of the participants act as students. Figure adapted from \cite{roleplayingtalk}.}
	\label{schematic}
	\end{center}
\end{figure*}

\subsubsection{Lab Simulation}\label{labsim}
The Lab Simulation is a roleplaying exercise in which the GTAs take turns acting as ``TA'' and as ``Students'' in the lab. Participants are arranged in pairs, and each pair facilitates the lab as TAs for 10 minutes while everyone else are the Students doing the lab experiments. While the TAs are facilitating, each of them is shadowed by an instructor who takes notes and then provides the TA with feedback. When their turn is over, the two people who had been TAs sit at a lab station and become Students, and another pair of participants become the TAs.

The lab room contains several setups for four different introductory physics experiments. To control logistics, each participant is assigned an experiment for which they will be TA and an experiment for which they will be Student. Lab materials are provided in advance, and participants are responsible for familiarizing themselves with the lab for which they will be TA. The left panel in Figure~\ref{schematic} shows a schematic of the classroom setting for the Lab Simulation.

To make things more interesting, roughly a third of the participants are contacted in private by the course instructor before the Lab Simulation, and are asked to \textit{sabotage} their experiments. For example, one participant may be asked to bring their laptop and refuse to work with their assigned lab partner; another participant may be asked to play with their phone instead of working on the lab. The participants tagged for sabotage are sworn to secrecy, and the sabotage is revealed only at the end of the activity, to everyone's delight.

\subsubsection{Microteaching}
Microteaching is the last three-hour session of the Orientation, where GTAs practice teaching in front of a group for the first time \cite{maslach}. Depending on the number of GTAs, Microteaching could last anywhere from one to three days, although each GTA only attends one session.

At the end of the Teaching Physics lesson (see Section~\ref{js2}), GTAs are presented with several introductory physics problems from which to choose. Each person selects one problem and signs up for one Microteaching session. The GTA is responsible for solving the problem and preparing to present and facilitate it. Solutions are not provided to the GTAs until after the activity is over, so they can get some practice refamiliarizing themselves with introductory physics concepts and problem solving techniques.

Each Microteaching session hosts a maximum of 10 GTAs. Each person has 10 minutes to facilitate their selected problem. We emphasize to the GTAs that they are not supposed to lecture; instead they must engage the audience (i.e., their peers, who act as students) by asking them questions and facilitating the solution process.

While a GTA facilitates, their peers are split into two groups -- see the right panel of Figure~\ref{schematic} for a schematic. When the facilitator is done (or 10 minutes have passed, whichever comes first), the course instructor and each group of peers fill out a Microteaching Feedback Form. Thus, each GTA ends up with feedback from three different sources. They will then use that feedback to write a Microteaching Debrief Essay.

\subsection{Follow-Up Meetings}
The Follow-Up Meetings happen during the semester, after the GTAs' teaching duties have begun. These are 50-minute class meetings taking place roughly every 2-3 weeks.

\subsubsection{Grading}
The Grading lesson is split into separate meetings for each type of teaching assignment, with each GTA attending the meeting that corresponds to their assigned teaching. Each meeting includes information about rubrics, a discussion of the grading rubric they are to follow, a Grading Flowchart to help GTAs streamline their grading workflow, and a thorough grading practice using real student solutions. An additional third meeting requires all GTAs to attend so they may learn how to use the software for electronic grading \footnote{Gradescope, \url{https://www.gradescope.com}}. All solutions and old exams that are used in the grading practice are stripped of student names and other identifying information before being made available to the GTAs.

\subsubsection{Midterm Evaluations and Time Management}
In this split-topic session, we first give GTAs a short questionnaire that serves as mid-semester evaluation of the course. Once they are done, we then introduce the idea of mid-semester evaluations and how they differ from end-of-semester evaluations. We then assign the Midterm Evaluations Project, an assignment in which the GTAs will craft their own midterm evaluation questions, administer them to their students, and then write a report with their midterm evaluation results. The GTAs are given three weeks to do this assignment.

The second half of this lesson focuses on time management. We discuss procrastination, look over the Workload Survey data (Section~\ref{workloadsurveys}) to date, then try to identify where the time goes based on the number of hours in one week. Finally, GTAs are introduced to the Important/Urgent time management matrix \cite{matrix}, and we discuss tips and strategies for managing their time.

\subsubsection{Teaching Videos}
In this session, we present GTAs with several videos of other GTAs teaching. All the videos are clips from classroom observations done in previous years. For each video we then discuss what was happening in the video, what they think the GTA in the video did well, what they think the GTA in the video needs to improve on, and what they would do differently.

\subsubsection{Teaching and Research}
This session focuses on the transferable skills from teaching that are useful for academic and non-academic careers. We first start by asking the GTAs what skills they have developed or improved on during their teaching assignment this semester, then compare their answers to what their expectations were before the start of the semester. We introduce the research-validated idea that teaching experience can improve graduate students' research skills \cite{feldon}, and discuss online resources about physics Ph.D. employment \cite{hiring}. We then do an activity that takes up the majority of the class time. For this activity, the course instructor places printouts of job ads for a variety of academic and non-academic jobs on the whiteboards. Each GTA is then asked to pick two job ads, one academic and one non-academic, then we do a think-pair-share. Individually, each person reads their selected ads and identifies the transferable skills. In pairs, they compare and contrast job ads. Then everyone shares the transferable skills they identified, how they are related to the GTA job, and whether they appear in academic or non-academic job ads (or both). We then briefly talk about the differences between academic and non-academic jobs, and we close the lesson with a list of online resources for finding jobs.

\subsubsection{Concluding Remarks}\label{concrem}
In the last class meeting of the semester, we revisit the Workload Survey data to see how the GTAs are spending their time. Then we ask them how their first semester in graduate school has been so far, and we compare their answers to their expectations before the start of the semester. After this we discuss a couple of campus resources: the Career Development Roadmap \footnote{\url{https://grad.gatech.edu/career-roadmap}} and Tech to Teaching \footnote{\url{https://www.ctl.gatech.edu/content/tech-teaching-0}}. We close the session with some end-of-semester assessments, and an informal discussion of how the semester went for everyone.

\subsection{Out-of-Class Activities}
There are several activities that are part of the Physics GTA Preparation course but that do not happen during class meetings.

\subsubsection{Workload Surveys}\label{workloadsurveys}
Although this is not technically an ``activity,'' we include it under this category since it happens outside of class and it is not technically an ``assignment.'' At the end of each week, we send the GTAs a short survey (created with Google Forms) in which they indicate the amount of hours they have spent that week on their various GTA duties. A separate section of the survey also asks them to indicate the number of hours they have spent on their graduate coursework outside of going to class. The purpose of this activity is to determine if GTA duties are staying within the 12-13 hours/week time limits of the teaching assistantship, to identify the duties that take up the most time, and to find out how many hours per week the GTAs are spending on their own coursework. We discuss the results of the workload surveys a few times during the semester to determine if there are ways in which the GTAs' workflows can be made more efficient.

\begin{figure}[htp]
    \begin{center}
	\includegraphics[width=0.4\textwidth]{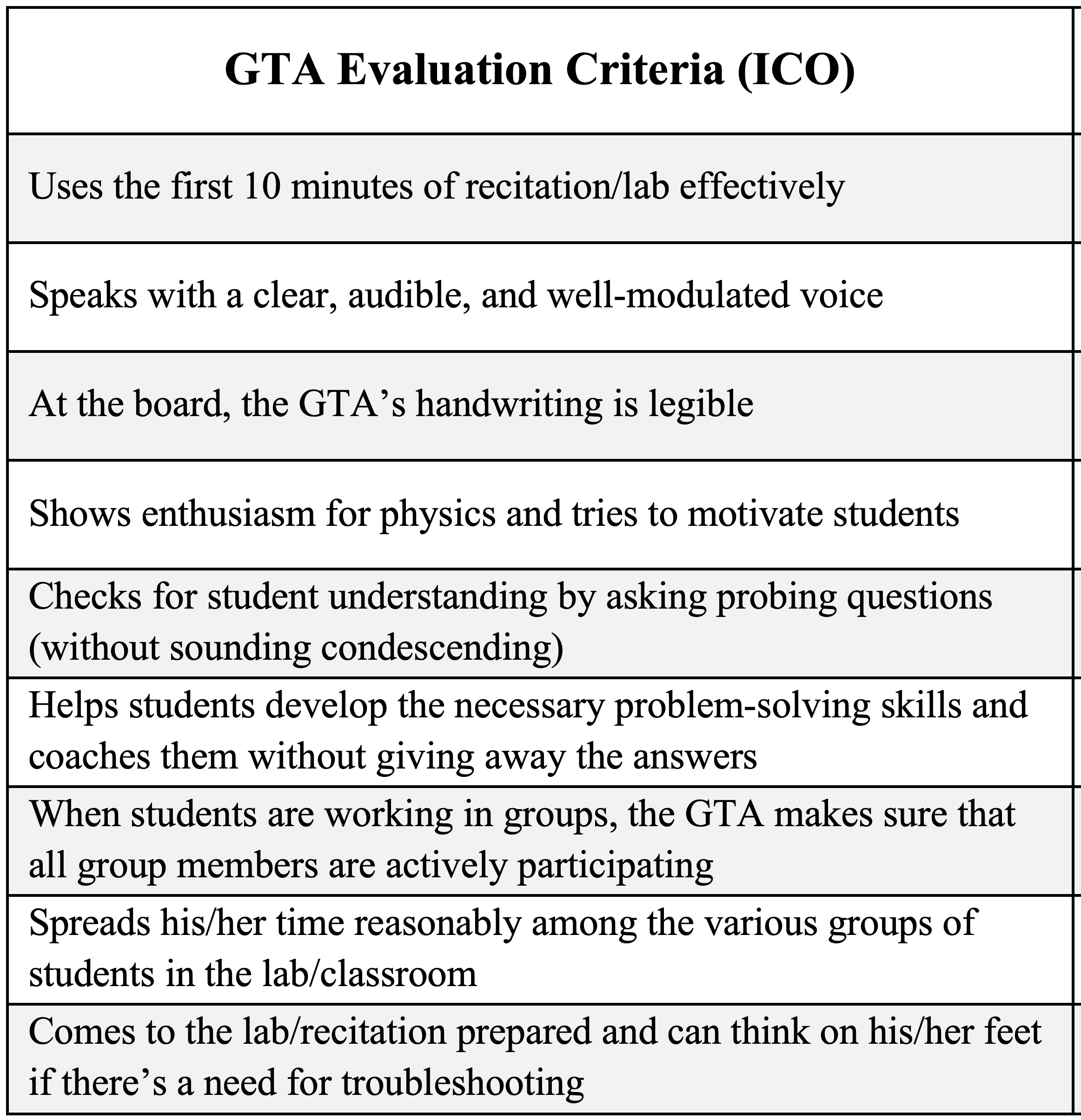}
	\caption{Evaluation criteria for the individual classroom observations (ICO). Each criterion is evaluated in a four-point scale: Excellent, Good, Just OK, Need to Improve.}
	\label{ico}
	\end{center}
\end{figure}

\subsubsection{Classroom Observations}
Each GTA is observed twice during their first semester of teaching. The first observation happens early in the semester (early September) and the second observation is later in the semester (late October). The observations are scheduled after the first week of GTA duties via a Google Form, where the GTAs indicate their preference for what section they would like to be observed in. In the form, the GTAs are also asked to list one aspect of teaching in which they want additional feedback during each observation. If the GTA gives permission, the observations are video recorded. The video is only shared with the person being recorded, and at the end of the semester anyone can request that their video(s) be deleted if they so wish. 
Since it is logistically difficult to observe all first-time GTAs during the entirety of one lab or recitation section (particularly those labs that last three hours), we limit the observations to 30 minutes. At the start of the observation, the observer and the GTA let the students know the purpose of the recording and ask if any students prefer to not be included in the video. If any students answer positively, the observer skips their table when video recording. If the observation starts at the beginning of lab/recitation, the observer records the GTA's introduction, which should last 10 minutes or less. Throughout the observation, the observer follows the GTA around with either a video camera or a clipboard, recording the GTA-student interactions and making note of how the GTA performs according to the rubric reproduced in Figure~\ref{ico}.

By the time of this writing, we have accumulated approximately half a terabyte of video data from classroom observations. These videos are used to provide each GTA with almost immediate feedback for improvement (the filled-out rubric, with detailed written feedback, is emailed to each GTA within two weeks of each observation). GTAs are encouraged to be constantly active during the lab/recitation, visiting groups and checking on their progress, and assisting students with guiding questions when they ask for help. Future work will include analysis of the videos to characterize the interactions between GTAs and students (e.g., who initiates, the length of the interaction, what happens during the interaction \cite{riot1,scherr,stang}).

\subsubsection{Mentoring Meetings}
Starting in 2017, the more senior graduate students have conducted peer mentoring sessions with the first-year graduate students. We have incorporated these as part of the GTA preparation class since the majority of first-year graduate students are GTAs as well. Three mentoring session happen each semester: the first one about academics, the second one about guidance and support, and the third one about career options. The peer mentors are older graduate students at various stages in their careers.

\section{Evolution of the Curriculum}
\begin{figure*}[ht]
    \begin{center}
	\includegraphics[width=\textwidth]{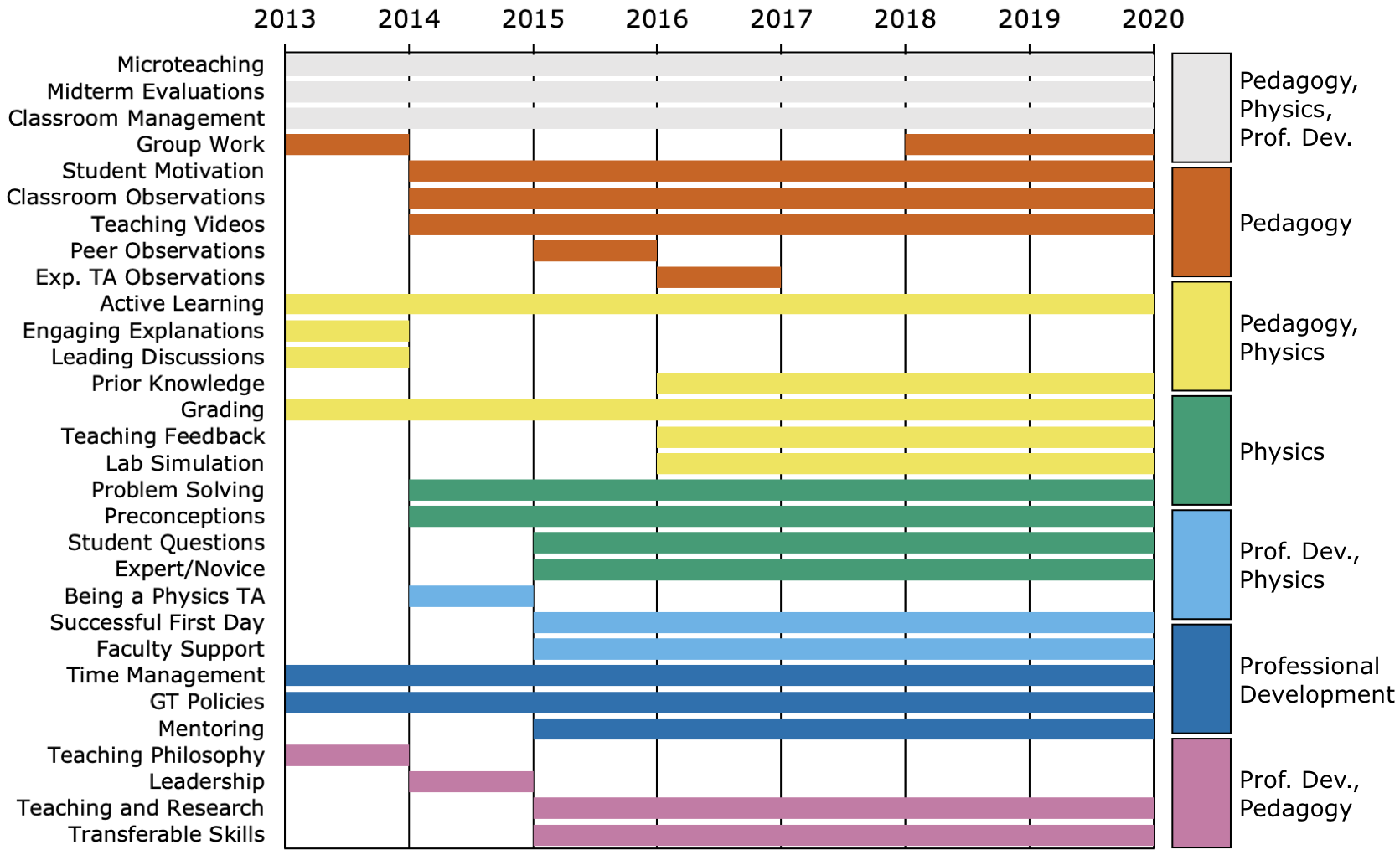}
	\caption{Evolution of the curriculum over the years, from 2013 to 2019. Each class module, topic, or activity is color-coded according to how it maps onto the 3P Framework. Note that some course elements have been there since the beginning, others came and went, and others still started later and have since remained. In the end, the present-day curriculum is more robust and comprehensive than when the course first started.}
	\label{evolution}
	\end{center}
\end{figure*}

How did the course evolve from the left panel to the right panel in Figure~\ref{firstlast}? A timeline diagram of the curriculum evolution is shown in Figure~\ref{evolution}. Each topic is color-coded according to how it maps onto the 3P Framework. The early gaps in the curriculum are clearly visible here, and we can also see the course become more comprehensive over the years.

Instead of chronologically explaining all the changes we made each year, we will focus on the course elements that have persisted since the beginning, the elements that failed and were eliminated quickly (``false starts''), and the elements that were added later and have proven successful. Finally, we will briefly describe the changes that were necessitated in Fall 2020 by the COVID-19 pandemic and the move to remote instruction.

\subsection{Persistent Over the Years}
\textit{Microteaching} is the first opportunity the new GTAs have to practice teaching in front of an audience and to receive feedback on their performance \cite{maslach}. In our experience and that of other researchers \cite[e.g.,][]{hume}, first-time GTAs consider it to be a highly valuable and useful experience. The Physics GTA Preparation course has included a Microteaching activity from the very beginning; however, the activity itself has not remained static. In 2013 and 2014, GTAs were given a list of physics topics, and each person would pick a topic and prepare a 10 minute lesson on their selected topic. Although GTAs in general considered Microteaching to be very useful, some indicated that they would prefer to have more guidance on what they are supposed to present while microteaching. Thus in 2015 we changed the format. Instead of selecting a topic, each GTA would select an introductory physics problem that they would then facilitate for their peers. The new format continues to this day.

\textit{Midterm Evaluations} is a dual-purpose lesson. First, the GTAs provide the instructor with feedback on the GTA preparation course, and then the GTAs are assigned a project in which they have to write their own midterm evaluations and administer them to their students to receive feedback from them, then write a report about their results. This is another lesson that has existed since the start, and for this one the format has not changed much over the years. The only difference is that over the years we have collected midterm evaluation questions from GTAs, and now we provide the new GTAs with sample midterm evaluations based on the work done by previous first-time GTAs. The GTAs are now given the opportunity to write their own questions, or mix-and-match questions from the samples, or copy a sample midterm evaluation wholesale, as long as they explain their reasoning for using those questions and the type of feedback they sought from their students.

\textit{Classroom Management} is another session title that has existed since the beginning, though the contents have changed over the years. It started as a short session with tips to manage a classroom and discussion of a handful of case studies. Group Work, which was a session on its own in 2013, was absorbed into Classroom Management in 2014. Between 2014 and 2017, Group Work was discussed in a very minimal way during Classroom Management, in the context of one single case study. However, we realized that GTAs were having issues managing groups of students, so we re-expanded Group Work while keeping it within the Classroom Management session. This session now also includes information about what to do on the first day of class and how to motivate students.

\textit{Active Learning} was originally a separate session which only happened in 2013. The concept of active learning, however, has remained an integral part of the class, particularly in the Teaching Physics lesson (Section~\ref{js2}), since we want GTAs to teach their labs and recitations by interactively engaging with the students. We now also provide the GTAs with references from PER \cite[e.g.,][]{hake} to emphasize how important and effective this style of teaching is. Another difference is that in the first two years of the course, we included a discussion of Bloom's Taxonomy \cite{blooms}. We have since removed this discussion given that it is not essential for the GTAs as novice instructors.

\textit{Grading} is a necessary part of any GTA preparation course, especially when a majority of the first-time GTAs have no prior teaching experience. From the beginning, the lesson started by discussing rubrics. It is interesting to note that many GTAs in the early years of the course said they had never heard of rubrics, but in later years most GTAs report knowing what a rubric is. The few changes that have been made to the Grading lesson have been necessitated by the different types of grading done in each GTA assignment, with each type of assignment having its own Grading session now.

\textit{Time Management} is a very important skill that will help GTAs not just in performing their teaching duties but also in their coursework and in their future careers. From the beginning, the Time Management lesson has introduced GTAs to the Eisenhower Important/Urgent Principle \cite{matrix}, in which tasks are categorized according to whether or not they are important and whether or not they are urgent. In later years we have added case studies and a spreadsheet for GTAs to list how many hours each of their weekly tasks require. Additionally, in 2015 we began assigning weekly Workload Surveys for GTAs to reflect on the time spent on their different teaching tasks.

\textit{Georgia Tech Policies} is a topic that needs to be covered since GTAs are employees of the Institute. There are four policies that are always discussed: Academic Integrity, the Office of Disability Services, Sexual Misconduct, and FERPA; additionally, we make sure to emphasize the importance of the Institute's Policy of Nondiscrimination. The lesson always includes case studies to discuss the nuances of each policy. In 2017 we created the OK/NOT-OK Game to further discuss the policies in a fun and engaging manner. The game was an instant hit. The GTAs were deeply engaged with the game, lots of laughs were had at the obvious scenarios, spirited discussions sprouted around the non-obvious scenarios, and everyone in general (including the course instructor) seemed to have a great deal of fun.

\subsection{False Starts}
\textit{Engaging Explanations} was a separate lesson in the first year of the course that was streamlined and absorbed into the Teaching Physics lesson (see Section~\ref{js2}).

\textit{Leading Discussions} was a session in which we talked about strategies for a GTA to lead a discussion, and how to respond to questions for which they do not know the answer. This session only existed in the first year of the course. Feedback from the GTAs indicated that they felt it was useless for their actual teaching assignments, and that we spent too much time on it that could have been spent on more hands-on practical activities. The session was thus eliminated, and the topic of what to do if you do not know the answer to a question was absorbed into Classroom Management.

\textit{Teaching Philosophy} was the way in which we injected professional development into the pilot semester of the course. The GTAs said it was not useful for them since most of them planned on careers in industry. In the second year of the course we replaced Teaching Philosophy with \textit{Leadership}, in which we discussed styles of leadership and how to develop leadership skills. The GTAs that year did not consider this useful either, so it was also eliminated.

\textit{Being a Physics TA} was a big portion of the Introduction and Policies session in 2014, and it was designed to help GTAs develop their identity as educators. It was not well-received because GTAs felt it was ``preachy.'' We eliminated this as an explicit discussion and instead focused on developing activities within the other course modules to help GTAs develop their professional identity.

\textit{Peer Observations} was an activity we included in 2015, in which GTAs would observe each other in groups of three and gave each other feedback. This activity received a lukewarm response, with some people enjoying it and other people hating it. Some GTAs said that they did not feel qualified to give useful feedback to their peers, while others said their peers were not qualified to give them useful feedback. In 2016 we replaced it with \textit{Experienced TA Observations}, in which each first-time GTA was assigned to observe an experienced GTA who was also teaching the same class that semester. This activity encountered severe logistics difficulties because on that particular semester there were very few experienced GTAs teaching the introductory physics classes. It is safe to say the activity was a disaster, and we quickly eliminated it from the curriculum.

\subsection{Newer and Successful}
Several new topics were added to the class in its second year -- \textit{Student Motivation} (included as part of the Classroom Management module), \textit{Classroom Observations} (in which GTAs are observed by an instructor and given feedback for reflection and improvement), \textit{Teaching Videos} (in which GTAs watch videos of experienced GTAs and critique them), and \textit{Problem Solving} and \textit{Preconceptions} (in which GTAs participate in activities to help them facilitate problem solving and identifying student preconceptions). All of these topics have remained largely unchanged over the years, with very few subtle improvements. For example, originally each GTA was only observed once per semester; now they are observed twice, once in early September and again in late October. In the present-day Teaching Videos lesson the GTAs watch videos from past years' classroom observations, so the GTAs they are watching and critiquing are first-time GTAs instead of experienced GTAs. The Problem Solving and Preconceptions topics are part of the Teaching Physics module (see Section~\ref{js2}), and the only changes they have experienced over the years is the inclusion of new problems for the GTAs to solve.

\textit{Anticipating Student Questions}, \textit{Experts and Novices}, and \textit{Prior Knowledge} were all added to the Teaching Physics module, making it even more comprehensive, to the point that it is now among the top-three most useful course topics according to the GTAs.

\textit{Strategies for having a successful first day of class} were added to the Classroom Management module, to ensure that GTAs would start their teaching on the right foot. In 2018 we also included an assignment to reflect on the first week of teaching. This way GTAs keep in mind the things that went well and identify the things that could have gone better and that they can improve on.

\textit{Faculty Support} is something the GTAs requested time and time again. Between 2015 and 2017, we had a faculty guest speaker come into the Introduction and Policies lesson to talk to the GTAs about teaching and professional development. We have not been able to have a guest speaker in the last three years because of scheduling conflicts; however, the faculty who are in charge of supervising the GTAs (the coordinators for the introductory physics classes) are fully on-board with the class and have provided the instructor with valuable resources for the GTAs.

\textit{Mentoring} is something very important for the professional development of first-time GTAs. We initially included mentoring by creating an unstructured lesson in which GTAs were welcome to ask questions and discuss any issues or difficulties they were having. The unstructured nature of the lesson was not well-received, so we eliminated it after one year. The following year we allowed a group of senior graduate students to address the GTAs in peer mentoring. This kind of mentoring was somewhat unstructured at first, but from 2017 onward it has been well-structured into three peer mentoring sessions covering topics about academics, guidance and support, and career options. This allows the first-time GTAs to meet some of the senior graduate students in the department and learn from their experiences.

After the failure of Teaching Philosophy and Leadership, we found ourselves wondering how to include more explicit professional development into the course. The answer came to us by thinking about \textit{transferable skills} that can be developed while teaching and that can be useful for a physicist even outside of the classroom. To this end we created the \textit{Teaching and Research} module, in which GTAs compare academic and non-academic job ads and identify the transferable skills required for each.

The most recent addition to the class is the \textit{Lab Simulation}, which is similar to Microteaching, but in a lab environment. The activity is a fun roleplaying exercise in which GTAs take turns being ``TAs'' and ``Students'' (Section~\ref{labsim}).

\subsection{COVID-19 and Remote Instruction}\label{covid}
The COVID-19 pandemic necessitated the move to online/remote instruction in Fall 2020. As such, the GTA preparation course had to be modified in terms of content and delivery to adjust to the circumstances. 

We attempted to keep the curriculum as close as possible to the Fall 2019 iteration, including the three-part structure of Orientation, Follow-Up Meetings, and Out-of-Class Activities. 

The meetings were not in-person but rather through videoconferencing. During the Orientation meetings, the GTAs were asked to keep their video camera on so that everyone could get to know everyone else. In the Follow-Up meetings, GTAs were given the freedom to choose whether to turn on their cameras or not, which resulted in about two thirds of them preferring to keep their video off.

Group activities were done by randomly assigning GTAs to breakout rooms. This allowed for small-group discussions, though the nature of the breakout room technology prevented the course instructor from observing the entire class at once. Mentoring meetings with senior graduate students also made use of the breakout room technology. Activities like the OK/NOT-OK game were conducted through student response systems (clickers). Activities that normally made use of the walls of the classroom (such as the Teaching and Research jobs activity) were instead done through online forms. The Lab Simulation had to be eliminated, as there was no way to do the activity in person without violating social distancing guidelines. Classroom Observations did not happen due to logistics and time constraints.

GTAs signed up for Microteaching through an online form, and submitted their feedback to their peers through a different online form. Once all the GTAs arrived at the video conference, they were split at random into two breakout rooms. The person selected to facilitate was sent to Room 1, where they would introduce the problem for their peers (who were acting the role of students). After five minutes, the facilitator was moved into Room 2, where they had to check the work that the 'students' had done while the facilitator was with the other group. This simulated the experience they would have teaching labs and recitations during the semester. The \textit{sabotage} element of the Lab Simulation was brought into the online Microteaching, to prepare them for possible worst-case scenarios. All GTAs were told that people in Room 2 would be working on the problems while the facilitator taught in Room 1. However, the people in Room 2 were instructed to not work on the problem and display ``bad student behaviors'' to the facilitators, such as not answering questions, not wanting to work on the problem, trying to dominate the discussions, etc. This resulted in a more fun activity, and the GTAs were able to practice what to do if faced with such a situation while teaching.

\section{Discussion}\label{currdevdisc}
For many years, the training of first-time GTAs in the School of Physics at Georgia Tech lacked cohesion and continuity. In a partnership with the Center for Teaching and Learning, we developed a Physics GTA Preparation course that would prepare and motivate GTAs for teaching. Teaching the course for the first time allowed us to identify the aspects of the course that needed changing to make it more relevant, useful, and valuable to our GTAs.

We developed the 3P Framework to better visualize the course contents. With the 3P Framework we postulate that in order to have a comprehensive GTA preparation program there must be a full integration between pedagogical knowledge, physics content, and professional development strategies. Under the guidance of the 3P Framework, we revised the course yearly and implemented changes to improve the curriculum. The 3P Framework can be used in other institutional or disciplinary contexts as a guideline for how to develop a comprehensive GTA preparation program.

Some elements of the course have been present from the start, such as Microteaching, Midterm Evaluations, Grading, and Time Management. But even these have gone through changes, some subtle and some drastic, to make them better and more useful year after year. Other elements were unfortunately awful failures that lasted only one year. Nevertheless, we persisted with the yearly revisions and developed new activities that have stood the test of time. The COVID-19 pandemic necessitated further changes to adjust the content and delivery of the course to the new reality of remote instruction.

By the time of this writing, the Physics GTA Preparation course is a well-established, comprehensive, stable, and long-running professional development program for first-time GTAs in the School of Physics. The course been effective at improving GTAs' self-efficacy and approaches to teaching (assessment details are provided in a companion paper, in preparation). The course has also been mostly well-received by the GTAs themselves -- see, for example, the following quotes taken between 2015 and 2018:

\textit{``Simply how it helped build my confidence in teaching students. I was SO nervous at first, but once we did the microteaching and labsim and I could see what my peers were doing, I felt way better.''}

\textit{``I loved the awareness and reflection this class brought to my teaching. I would not have grown as a TA nearly as much without this class.''}

\textit{``It was an eye-opener. A lot of mistakes were avoided because of this class.''}

\textit{``I would have been a thoroughly mediocre TA had it not been for this course. Thank you!''}

Of course, not everything has been sunshine and rainbows. We consider it almost tradition by now to get one extremely negative comment each year. However, our yearly review of GTAs' comments and feedback give us the impression that there are significantly more positives than negatives. 

\vspace{14pt}

\begin{acknowledgments}
We would like to thank Ed Greco for valuable discussions and feedback about GTAs and teaching labs; Eric Murray, Martin Jarrio, and Nicholas Darnton for information about the specifics of the GTA assignments they supervise; Elaine Rhoades, Danielle Skinner, Benjam\'{i}n Loewe, and Logan Kageorge for their assistance with Classroom Observations, Microteaching, and Lab Simulation; Andrea Welsh for her advice about issues concerning LGBT+ students and suggestions for scenarios for the OK/NOT-OK game; Kate Williams and Felicia Turner from the Center for Teaching and Learning, for their support with coordination and logistics for the Orientation sessions; and last but not least, we thank all the graduate students who have participated in our GTA preparation course since it was first established eight years ago.
\end{acknowledgments}


\bibliographystyle{apsrev4-1}  	
\bibliography{forarXiv}  	

\end{document}